\documentclass{article}

\usepackage{arxiv}

\usepackage[utf8]{inputenc} 
\usepackage[T1]{fontenc}    
\usepackage{hyperref}       
\usepackage{url}            
\usepackage{booktabs}       
\usepackage{amsfonts,amsmath,amssymb}       
\usepackage{nicefrac}       
\usepackage{microtype,caption,float}      
\usepackage{lipsum,bm}		
\usepackage{graphicx}
\usepackage{natbib}
\usepackage{doi}

\newcommand{\Xvec}{\bm{X}_i}

\newcommand{\bvec}{\bm{b}_i}

\newcommand{\Xmat}{\bm{X}_{ij}}
\newcommand{\Zmat}{\bm{Z}_{ij}}
\newcommand{\delvec}{\bm{\gamma}}
\newcommand{\betvec}{\bm{\beta}}
\newcommand{\Dset}{\mathcal{D}_i}
\newcommand{\taumat}{\bm{\Lambda}}

\newcommand{\fdi}{F_i(\Dset)}
\newcommand{\fld}{f(d)}
\newcommand{\fldi}{f_i(d)}

\newcommand{\tauvec}{\bm{\tau}}

\title{Heterogeneous Effects in the Built Environment}

\author{ \href{https://orcid.org/0000-0001-7071-7873}{\includegraphics[scale=0.06]{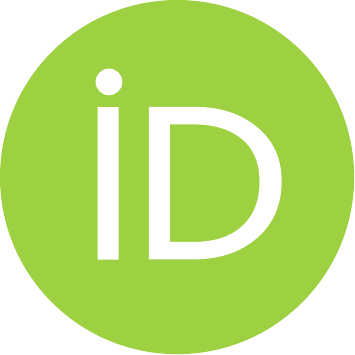}\hspace{1mm}Adam T. Peterson} \\
	Department of Biostatistics\\
	University of Michigan\\
	Ann Arbor, MI 48104 \\
	\texttt{atpvyc@umich.edu} \\
	\And
	{Brisa N. S\'anchez} \\
	Department of Biostatistics and Epidemiology\\
	Drexel University\\
	Philadelphia, PA \\
	\texttt{bns48@drexel.edu} \\
	 \AND
	 Emma Sanchez-Vaznaugh \\
	 Department of Public Health\\
	University of California San-Francisco \\
	San Francisco, CA
}

\date{}

\renewcommand{\shorttitle}{\textit{arXiv}}

\hypersetup{
pdftitle={A template for the arxiv style},
pdfsubject={q-bio.NC, q-bio.QM},
pdfauthor={David S.~Hippocampus, Elias D.~Striatum},
pdfkeywords={First keyword, Second keyword, More},
}

\begin{document}
\maketitle

\begin{abstract}
We present an approach to estimate distance-dependent heterogeneous associations between point-referenced exposures to built environment characteristics and health outcomes. By estimating associations that depend non-linearly on distance between subjects and point-referenced exposures, this method addresses the modifiable area-unit problem that is pervasive in the built environment literature. Additionally, by estimating heterogeneous effects, the method also addresses the uncertain geographic context problem. The key innovation of our method is to combine ideas from the non-parametric function estimation literature and the Bayesian Dirichlet process literature. The former is used to estimate nonlinear associations between subject's outcomes and proximate built environment features, and the latter identifies clusters within the population that have different effects. We study this method in simulations and apply our model to study heterogeneity in the association between fast food restaurant availability and weight status of children attending schools in Los Angeles, California.
\end{abstract}


\section{Introduction}
\label{sec:Intro}

The relationship between amenities in or near residential, work or school–neighborhood environments and health is receiving increasing attention, given that these environments can influence health-related behaviors and subsequent outcomes.
Where spatial proximity to supermarkets is associated with diet, so too are recreational facilities associated with physical activity and fast food restaurants near schools associated with child obesity \citep{baek2016distributed,baek2017methods,kaufman2019neighborhood,kern2017neighborhood}.
Work in this area has been limited by the lack of knowledge of what geographic units are most relevant for exposure assessment, i.e. the well known modifiable unit areal problem (MAUP) \citep{fotheringham1991modifiable,spielman2009spatial,wong2009modifiable,guo2004modifiable,ji2009spatial,james2014effects}. 
Additionally, there may also be measured or unmeasured person-level behaviors or characteristics that give rise to the ``uncertain geographic context problem'' (UGCP) \citep{MACINTYRE2002125,kwan2013beyond,kwan2018limits}. 
Whereas the former establishes that using different spatial units or spatial scales to define exposure measures will yield different estimates of association, the latter acknowledges that the most relevant spatial unit may differ from place to place or subject to subject due to place or person characteristics such as predominant transport modes in a given area or vehicle ownership, among others. 
\par 
Recent work has begun addressing these issues by foregoing the pre-specification of the spatial unit used to construct exposure metrics \citep{baek2016distributed,peterson2018rstap}. Instead, the association between proximity to amenities of interest, broadly referred to as built environment features (BEFs), and subjects' outcomes is estimated as a continuous function of distance between subjects and amenities.  Whereas typical models regress the outcome on a BEF metric that depends on a pre-defined scale, these new methods use all the pair-wise distances between subjects and BEFs as inputs to the model. Specifically, 
in order to address the MAUP, an idealized smooth function $f(d)$ is used to represent the association between the health outcome of interest and a single BEF placed at distance $d>0$ from the subject.
Having $f(d)$ as the objective of inference enables the visualization of whether and how the association between availability of amenities and outcomes dissipates with distance, as well as estimation of the spatial scale, defined as the distance at which the association is negligible, i.e. $d: f(d) =0$.
\par 
The function $f(d)$ has been modeled in different ways:
 \cite{baek2016distributed} estimated $f(d)$ non-parametrically by first discretizing the distances into a grid, and using the count of distances within bins defined by the grid as predictors in a Distributed Lag Model (DLM), i.e., the count of distances within each bin are conceptualized as distributed lag predictors, indexed by the corresponding value of the grid. The coefficients corresponding to each distributed lag predictor are smoothed using splines, yielding estimates of $f(d)$ at the values of $d$ used to construct the grid. Alternatively, \cite{peterson2021spatial} modeled $f(d)$ parametrically, typically using exponential functions to enforce the substantive belief that the association between health outcomes and spatial availability of amenities monotonically decays across distance, e.g. $f(d) \propto \exp(-\frac{d}{\theta})$. 
However, the estimation of $f(d)$ at the population level, as the previous methods propose, fails to account for the concerns the UGCP raises regarding unmeasured person-level behaviors or place-level factors that may determine subject – or location – specific spatial association. 
\par Building upon their work in DLMs, \cite{baek2016hierarchical,baek2017methods} constructed a hierarchical DLM (HDLM) allowing for the estimated $f(d)$ to vary  between  subjects and or locations, according to pre-specified groups (e.g., different $f(d)$ by sex), as well as unexplained variation in the association (i.e., using the idea of random coefficients to estimate $f(d)$ for individual subjects).
However, the HDLM approach, has some disadvantages: (1) it uses discretized distances to estimate association across space, unnecessarily coarsening the exposure information; (2) it requires pre-specifying the groups where heterogeneity in the association may occur (covariates and or subjects); and (3) by enforcing that heterogeneity in the association estimates to occur at the subject-level through random effects, it loses possible gains in precision that could result from pooling subjects with similar levels of association.
\par 
Motivated by the desire to identify schools where pupils may be at greater risk of obesity related to the proximity of fast food restaurants (FFRs), we propose a model that clusters schools-specific association curves, $f(d)$ according to the strength of association between the spatial proximity of nearby FFRs and child obesity. Clustering provides investigators and policymakers with a greater understanding of the kinds of relationships that exist between students and their environment as well as identifies schools where students may be at greater risk, as 
identifying risk groups may help prioritize population level interventions.
The data for this motivating study consists of body weight status of children nested within schools across Los Angeles County during academic years 2001-2008.
 Distances between schools and FFRs are calculated from geocoded school addresses, supplied by the California Department of Education, and geocoded FFR business addresses from the National Establishment Time Series Database\citep{walls2013national}
 \par Our method uses the Dirichlet Process Mixture (DPM) prior and a spline basis function expansion to non-parametrically estimate both the number of cluster-BEF effects, and the nonlinear association functions across space, respectively. We name our method the Spatial Aggregated Predictor - Dirichlet Process, to reflect this dual non parametric estimation, but refer to it more generally as STAP-DP, given its potential for also modeling temporal exposures. Our approach is inspired by the work of \cite{rodriguez2014functional} and \cite{ray2006functional} on clustering functions using DPM family priors. 
 We use the penalized spline approach developed by \cite{o1986statistical} and further popularized by \citep{wahba1990spline,mgcv_wood} to construct the estimates of the association functions, and use the DPM to cluster the spline coefficients. 
\par Section 2 describes the model that estimates  homo- and heterogeneous BEF effects. 
 Section 3 studies the performance of the STAP-DP model in a variety of simulated data settings and discusses how the results may inform normative practice.
 Section 4 describes the application of the STAP-DP model to the motivating study on child obesityin Los Angeles. 
 We conclude our work with a discussion of the model and future directions to explore.

\section{Model}
\label{sec:Model}

We now introduce the STAP-DP framework, describing how we incorporate the estimation of heterogeneous BEF effects into a regression framework. We limit our discussion to the estimation of only one BEF's effects in space, FFRs for example, as the extension to multiple BEFs is straightforward. We organize our discussion into four parts. First, we build intuition for our approach by defining the STAP estimated via spline basis functions at the population level, i.e., homogeneous effect. Then, we define how to extend the STAP model to estimate heterogeneous effects -- at the latent cluster level --  for a univariate outcome. In the final two sections we generalize the clustering framework for repeated outcome measures and discuss estimation.

\subsection{The STAP Model}
\label{sec:Model_Univariate}
Suppose a continuous outcome $Y_i$ ($i=1,...N$) and corresponding covariates $\Xvec \in \mathbb{R}^{n \times p}$ are observed for a sample of $N$ subjects. Additionally, spatial data, $\Dset$ which contains distances, $d$, between subject $i$ and all FFRs within some substantively determined radius $R$, are also measured. The inferential objective is to estimate function $\fld$, which represents the expected difference in the outcome associated with placing a single FFR at distance $d$ after adjusting for covariates $\Xvec$. Defining $F(\mathcal{D}_i) := \sum_{d \in \mathcal{D}_i} f(d)$, as the aggregated FFR effect under the assumption of additivity, we complete the initial STAP model formulation:  

\begin{align*}
    Y_i &= \Xvec^{T}\delvec + F(\mathcal{D}_i) + \epsilon_i, \tag{1} \label{eqn:stap_general} \\
    \epsilon_i &\stackrel{iid}{\sim} N(0,\sigma^2),
\end{align*}

\noindent where $\epsilon_i$ is the residual error, with variance $\sigma^2$.

As mentioned in Section \ref{sec:Intro}, there are a number of approaches to model $f(d)$.  In this work, we propose to model $f(d)$ as a linear combination of basis functions, $\{\phi\}_{l=1}^L$, which allows us to rewrite $F(\mathcal{D}_i)$ as follows: 
\begin{align*}
    F(\mathcal{D}_i)&= \sum_{d \in \Dset} \fld  = \sum_{d \in \Dset} \sum_{l=1}^{L} \beta_l\phi_l(d), \label{eqn:f_def} \tag{2}
\end{align*}
where $\phi_l(d)$ is the evaluation of the distance through the $l$th basis function and $\beta_l$ is the corresponding regression coefficient. 
In this work we use $L$ spline basis functions defined across a set of equally spaced knots, though other knot placements or basis functions could be used. 
In order to avoid over fitting when $L$ is large, the regression coefficients are regularized through the use of a quadratic penalty on $\betvec$ implemented through a smoothing matrix $S$ and tuned by penalty parameter $\tau$. 
We use the difference penalty matrices of \cite{eilers1996flexible}, a widely used spline penalty formulation.
\par Within a Bayesian paradigm, this penalty is equivalent to  specifying a multivariate normal prior with improper precision matrix $\tau S$.
We adopt a variant of this Bayesian approach and, to improve computational efficiency in our more complex model formulations discussed in the next subsection, we first transform the spline basis function expansion matrix, $\Phi(d)$, such that the transformed coefficients can have independent normal priors \citep{wood2004stable,wood2016just}. While centering constraints are often imposed on $\Phi(d)$ to avoid collinearity with the intercept in $\bm{X}$, this constraint is not needed in our model (see supplementary material).
Given that $r_S =$rank($S)<L$, two precision parameters for the priors are used, one for the first $r_S$ coefficients and a second for the last $L-r_S$ coefficients:
\begin{align*}
    \betvec_1 &\sim MVN_{r_S} \left (\bm{0},\sigma^2 \tau_1^{-1} \bm{I}_{r_S}   \right ) \quad \betvec_2 \sim MVN_{L-r_S}(\bm{0},\sigma^2 \tau_2^{-1}\bm{I}_{L-r_S}) \tag{3} \label{eqn:prior}\\
    \tau_z &\stackrel{iid}{\sim} \text{Gamma}(a_{\tau},b_{\tau}) \quad z = 1,2.
\end{align*}
 In $(\ref{eqn:prior})$ we denote $\betvec_z$, $z=1,2$, as the regression coefficients in the penalty range and null space, respectively. Correspondingly, $\tau_1$ and $\tau_2$ are the respective precisions for these separate subsets of $\betvec$.
For ease of further exposition we define $\taumat$ as the diagonal covariance matrix which has $\tau_1^{-1}$ as the first $r_S$ diagonal elements and $\tau_2^{-1}$ as the last $L-r_S$ diagonal elements, so that the prior in (\ref{eqn:prior}) can be written simply as $\betvec \sim MVN_L(\bm{0},\sigma^2\taumat)$.
We place independent conjugate Gamma priors on $\tauvec= (\tau_1,\tau_2)$ so that both $\betvec$'s and $\bm{\tau}$'s conditional posterior distributions are available in closed form.

\subsection{STAP-DP with Univariate Outcomes} 
In alignment with this work's goal to estimate heterogeneous effects, we replace $F_i(\mathcal{D}_i)$ with $\fdi$ in (\ref{eqn:stap_general}) while allowing for clustering in the $\fldi$.  Given that $\fldi$ is represented by the fixed spline functions and random coefficients $\betvec$, we implement this clustering goal by  placing a DP prior on the vector of regression coefficients, $\betvec$, and associated penalty parameter, $\bm{\tau}$: 
\begin{align*}
    (\betvec,\bm{\tau}) &\sim P \label{eqn:stapdp1} \tag{4}\\
    P  &\sim DP(\alpha,P_0)\\
    P_0 &\equiv MVN_{L} \left(\bm{0},\sigma^2 \taumat \right )\times \prod_{z=1}^{2} \text{Gamma}(a_{\tau},b_{\tau}).
\end{align*}
In (\ref{eqn:stapdp1}), $P$ is a random measure drawn from Dirichlet Process $DP(\alpha,P_0)$, where $\alpha>0$ is a concentration parameter reflecting the variability of distribution $P$ around base measure $P_0$ \citep{ferguson1973bayesian,gelman2013bayesian}. 
$P_0$ is chosen to retain the prior previously discussed in (\ref{eqn:prior}).
\par By placing the DP prior on $(\betvec,\tauvec)$, clustering is induced on the $\fldi$ as can be seen from the stick breaking construction of the DP:  $P = \sum_{k=1}^{\infty} \pi_k \delta_{(\betvec^\star,\tauvec^\star)}(\cdot)$. In this representation $\pi_k$ represents the probability the $i$th observation is assigned to the $k$th exposure function and $\delta(\cdot)$ is the dirac-delta function. Each $\pi_k$, itself is composed of the ``broken sticks'' created from variables drawn from a Beta distribution: $\pi_k = v_k \prod_{u<k} (1-v_u); v_k \sim \text{Beta}(1,\alpha) $.
\par  Combining all these pieces together, our proposed STAP-DP model for univariate outcome $Y_i$ takes the following form:
\begin{align*}
    Y_i &= \Xvec\delvec + \sum_{d \in \Dset}  \sum_{l=1}^{L} \beta_{il}\phi_l(d) + \epsilon_i \label{eqn:mod1} \tag{5}\\
    \epsilon_i &\stackrel{iid}{\sim} N(0,\sigma^2)\\
    (\betvec,\tauvec) &\sim P\\
    P &\sim DP(\alpha,P_0) \\
    P_0 &\equiv MVN_L\left(\bm{0}, \sigma^2 \taumat \right )\times \prod_{z=1}^{2}\text{Gamma}(a_{\tau},b_{\tau}).
\end{align*}
 A final comment is warranted regarding the choice of the number and placement of the $L$ knots in constructing the splines. While our approach follows previous work in placing a sufficient number of knots equally across the domain of observed distances, deciding what number of knots is ``sufficient'' requires greater statistical judgement than in standard applications. 
 Clusters may be more difficult to detect when the dimension on which clusters are formed (i.e., number of coefficients) is large and the between-cluster differences are small (low signal effects).
 Conversely, more clusters may be identified in a setting with a stronger signal and greater number of knots.
 Thus, $L$ must be chosen to balance accuracy in both function estimation \textit{and} cluster discrimination.

\subsection{STAP-DP with Repeated Measurements}
\label{sec:Model_Multivariate}
Extending (\ref{eqn:mod1}) to correlated outcomes,
we consider the setting in which subjects are measured repeatedly over time, for $j=1,...,n_i$ occasions.
This results in outcome $Y_{ij}$ ($i=1,...,N,j=1,...,n_i)$ modeled as a function of covariates $\Xmat$, and their corresponding coefficients $\delvec$. The distance set adopts the new visit-specific index as well, i.e., $\mathcal{D}_{ij}$, indicating it may vary over time; for instance FFRs may open and close between measurement occasions. Finally, a subset of $\Xmat$, $\Zmat$, is included in the model, along with subject-specific coefficients $\bvec \sim MVN(0,\Sigma)$ to account for within subject variability in standard fashion \citep{fitzmaurice2008longitudinal}. Augmenting (\ref{eqn:stapdp1}) accordingly, we arrive at our final model:
\begin{align*}
    \tag{6} \label{eqn:long1}
    Y_{ij} &= \Xmat^{T}\delvec + \sum_{d\in\mathcal{D}_{ij}}\sum_{l=1}^{L}\beta_{il}\phi_l(d) + \Zmat^{T}\bvec + \epsilon_i\\
    \bvec & \stackrel{iid}{\sim} N(\bm{0},\Sigma)\\
    \epsilon_i &\stackrel{iid}{\sim} N(0,\sigma^2)\\
    (\betvec,\tauvec) &\sim P\\
    P &\sim DP(\alpha,P_0) \\
    P_0 &\equiv MVN_L \left ( \bm{0}, \sigma^2 \taumat \right  )\times \prod_{z=1}^{2}\text{Gamma}(a,b).
\end{align*}

\subsection{Estimation}
\label{sec:Model_Estimation}

In order to fit models of the form described in (\ref{eqn:stapdp1}) and (\ref{eqn:long1}), we truncate the DP so that a blocked Gibbs sampler  can be used to draw samples from the posterior \citep{gelman2013bayesian}. 
While this sampler is fairly straightforward, it bears mentioning that $\Phi(d)$ has to be adjusted at each iteration of sampling so that any DP components associated with 0 or some small number of observations are not included in the usual matrix inversion used to estimate the mean of the conditional posterior distribution for the regression coefficients, $\betvec^{*} = [\delvec,\betvec]^{T}$.
Instead, coefficients for those low-member cluster components are sampled with draws from the prior. For example, if on the $m$th iteration, none of the $N$ observations are assigned to the $k$th DP component, then the samples of the spline regression coefficients for that iteration, $\betvec_k^{(m)}$,  are drawn from a $MVN_L(\bm{0},\sigma^2\taumat_k)$ prior, where $\taumat_k$ is the cluster specific covariance matrix, and the columns of zeros that would otherwise be included in $\Phi(d)$ are omitted. 
\par  We present the closed form conditional posteriors and associated algorithm in the Supplementary Material. Our algorithm is implemented in C++ which can be called from our R package \texttt{rstapDP} \citep{rstapDP,supp}.\par For both our simulations and California data we use \texttt{rstapDP} to fit the STAP-DP in \texttt{R} (v.4.0.2) \cite{Rlanguage} on a MacOS Catalina operating system with a 2.8 GhZ Quad-Core Intel Core i7 processor.

\section{Simulations}
\label{sec:Simulations}
\subsection{Simulation Design}
For a given sample size, the ability of the STAP-DP model to correctly classify subjects depends on (a) the proportion of subjects belonging to that cluster, (b) the difference in the $f_i(d)$ functional forms, and (c) the distribution of distances (i.e., exposure information) present within each cluster.
As the first of these three principles follows straightforward sample-size intuitions, in this section we study the STAP-DP's ability to correctly recover cluster specific functions,$f_i(d)$, and cluster partitions in the latter two settings. Using simulated data we vary: (i) cluster effect size and (ii) distance distributions in order to see how these may impact correct cluster classification. 
We focus on evaluating cluster classification accuracy as it is the upstream predictor of all remaining model components, like the estimation of the $f_i(d)$, which are all standard Bayes estimators conditional on the correct cluster classification.
\par 
We evaluate our method's ability to correctly classify subjects using a partition loss function developed by \citep{binder1978bayesian} and used regularly in DP and other mixture model applications where label-switching may be of concern \citep{lau2007bayesian,wade2018bayesian,rodriguez2008nested}. Our employment of the loss function equally weights correct and incorrect classification, using the subjects' true and estimated class indicators, $\zeta_i,\hat{\zeta}_i$, respectively: 
\begin{align*}
  \tag{7} \mathcal{\psi}(\bm{\zeta},\hat{\bm{\zeta}}) = \sum_{(i,i'); i< i' < N} I(\zeta_i = \zeta_{i'},\hat{\zeta}_i \neq \hat{\zeta}_{i'}) + I( \zeta_i \neq \zeta_{i'},\hat{\zeta}_i = \hat{\zeta}_{i'}).\label{eqn:loss} 
\end{align*}
\noindent Conceptually, (\ref{eqn:loss}) tallies the number of times that observations $i$ and $i'$ are incorrectly assigned to different clusters, when they in fact belong in the same cluster, as well as tallies when they are incorrectly assigned to the same cluster.
\par In each simulation setting discussed below, we generate 25 datasets and then fit the STAP-DP model shown in (\ref{eqn:stapdp1}), truncating the DP at K= 50 and using weakly informative Gamma(1,1) priors on $\sigma^{-2},\alpha, \tau_1$ and $\tau_2$, respectively. We draw 2000 samples from the posterior distribution for inference via Gibbs Sampling using \texttt{rstapDP} after discarding 2000 initial samples for burn-in.
Across all 25 simulations we evaluate the loss (\ref{eqn:loss}) across all $M=2000$ iterations of the posterior samples drawn via Gibbs sampling. 
Given that the loss function does not have a standard range, we normalize the loss results by the maximum loss across all simulation settings, so as to make the results more interpretable relative to one another.  We have organized the files used to run the simulations in the STAPDPSimulations R package available via \href{ https://github.com/apeterson91/STAPDPSimulations}{Github}.

\subsection{Cluster Effect Size}
 Our first simulation study focuses on model performance as a function of the difference between two clusters' $f(d)$, defined in (8) below, with each observation having a 50\% probability of being assigned to either of these clusters. 
 The cluster function set-up is intended to mimic a hypothetical high and low risk population scenario, in which subjects with equivalent exposure to the same BEFs experience different effects according to which risk population they belong. 
 For each subject we generate a random number of distances uniformly so that the average number of BEFs is 15. Conditional on the number of distances, the distances themselves are then generated according to the ``Skew'' distribution shown in Figure S1 in the Supplementary Information. This distribution was selected in order to test our model's performance under a ``worst case scenario'', given that with this distribution there is relatively less exposure information at the distances where the BEF effects are non-zero.
 Specifically, the generative model takes the following form:
\begin{align*}
    Y_i &= 26 + .5Z_i + \sum_{d \in \mathcal{D}_i} f_{\zeta_i}(d) + \epsilon_i \tag{8} \label{eqn:sim}\\
    \epsilon_i &\sim N(0,\sigma^2=1) \qquad i = 1,...,200\\
    f_1(d) &=   \exp\left \{ \left (\frac{-d}{.5} \right)^5 \right\} \qquad  \\
    f_2(d) &= \nu \exp\left \{ \left (\frac{-d}{.5} \right)^5 \right\} \qquad \nu = (0,.25,.5,.75)\\
    P(\zeta_i=1) &=  P(\zeta_i=2)=.5; 
\end{align*}
where $Z_i$ is a covariate generated as a fair Bernoulli random variable, $\zeta_i$ is the subject specific cluster label indicating the true BEF effect, $f(d)$,  for the $i$th observation, and $\nu$ represents the varying effect size at $d=0$. 
\par
\begin{figure}[H]
    \centering
    \includegraphics[width = .6 
    \textwidth]{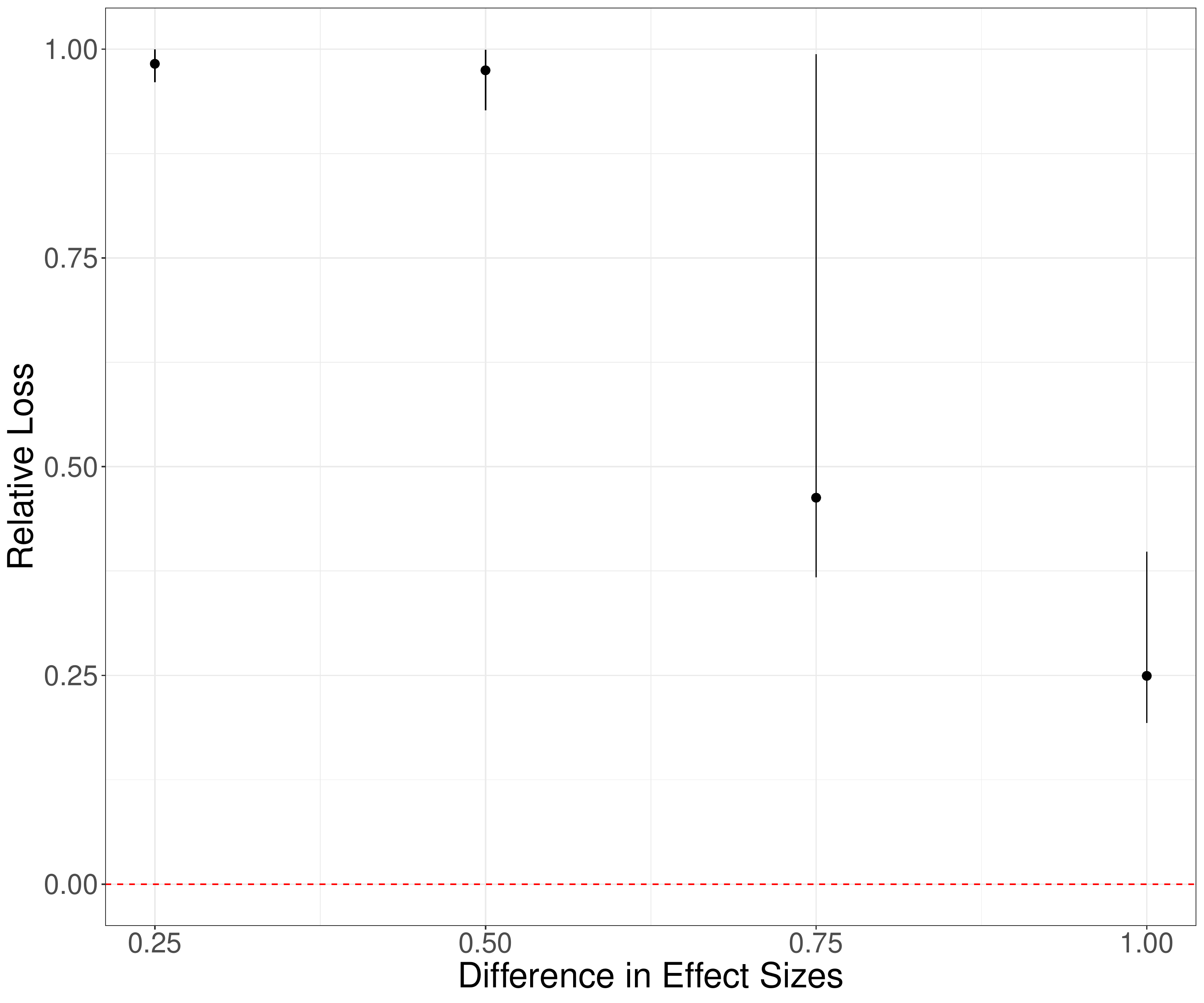}
    \caption{Relative loss as a function of the difference in effect size: $(1-\nu)$; see (8) for more details. Point estimates and error lines represent median, 2.5 and 97.5 quantiles of loss across simulations, respectively.}
    \label{fig:effect_loss}
\end{figure}

 The relative loss as a function of the effect size $\nu$ is shown in Figure \ref{fig:effect_loss}. As expected, one can see a decrease in relative loss and consequently improved classification as the effect size increases.

\subsection{Distance Distributions}

As our method non-parametrically estimates cluster functions $f_i(\mathcal{D}_i)$ across continuously measured space using a basis function expansion, correct estimation of the function requires there to be BEFs observed at the relevant distances, $d: f(d)\neq 0$, within the study area of interest. Of course these ``relevant'' distances are not known \textit{a priori} and so it is to the benefit of the investigator to err on the side of caution in specifying a larger study area if possible. 
However, despite any preparatory work that may be done to ensure an adequate area is included at the level of the sample study, it is not clear how differing distributions of distances at the latent cluster level may impact inference. For example, will suburbanites' lower exposure to proximate FFRs impact the ability of the stapDP model to discern the impact of FFRs on their health relative to their more exposure rich urban counterparts? For this reason our second simulation study examines how exposure to different distance distributions may impact classification.
\par 
We study this problem by considering three different generative distance distributions which we label ``Uniform'',``CA'' and ``Skew''.
The first, straightforwardly, refers to the idealistic - but unrealistic - scenario in which there is equivalent information available at all distances within the study area. 
The second two cases refer to more realistic situations in which there are more likely to be a higher number of BEFs found further away from the subject than close by – a consequence of area's quadratic growth as a function of distance. 
We create the first of these skewed distributions, ``CA'', by using maximum likelihood to fit a beta distribution to the distribution of distances in our motivating California data distance distribution and the second by altering a beta distribution to be a more extreme version of the first. 
We generate distances under each distribution for each cluster in order to examine how differing exposure patterns between clusters impact cluster classification. The densities of each of these distributions can be found in Figure S1 in the supplementary material.

Since the exposure information depends both on the distribution of distances and the number of BEFs, we generate scenarios where the amount of information increases as a function of the number of BEFs within the same distribution. We simulate data under the same model as proposed in (\ref{eqn:sim}), with $\nu=0.25$, to illustrate how a substantial, but not obvious, difference in cluster functions manifest across the varying distance distribution settings. Fitting our STAP-DP model under the priors and sampler settings previously described, we plot the results below in Figure \ref{fig:ddist_loss}.

\begin{figure}[H]
    \centering
    \includegraphics[width = \textwidth]{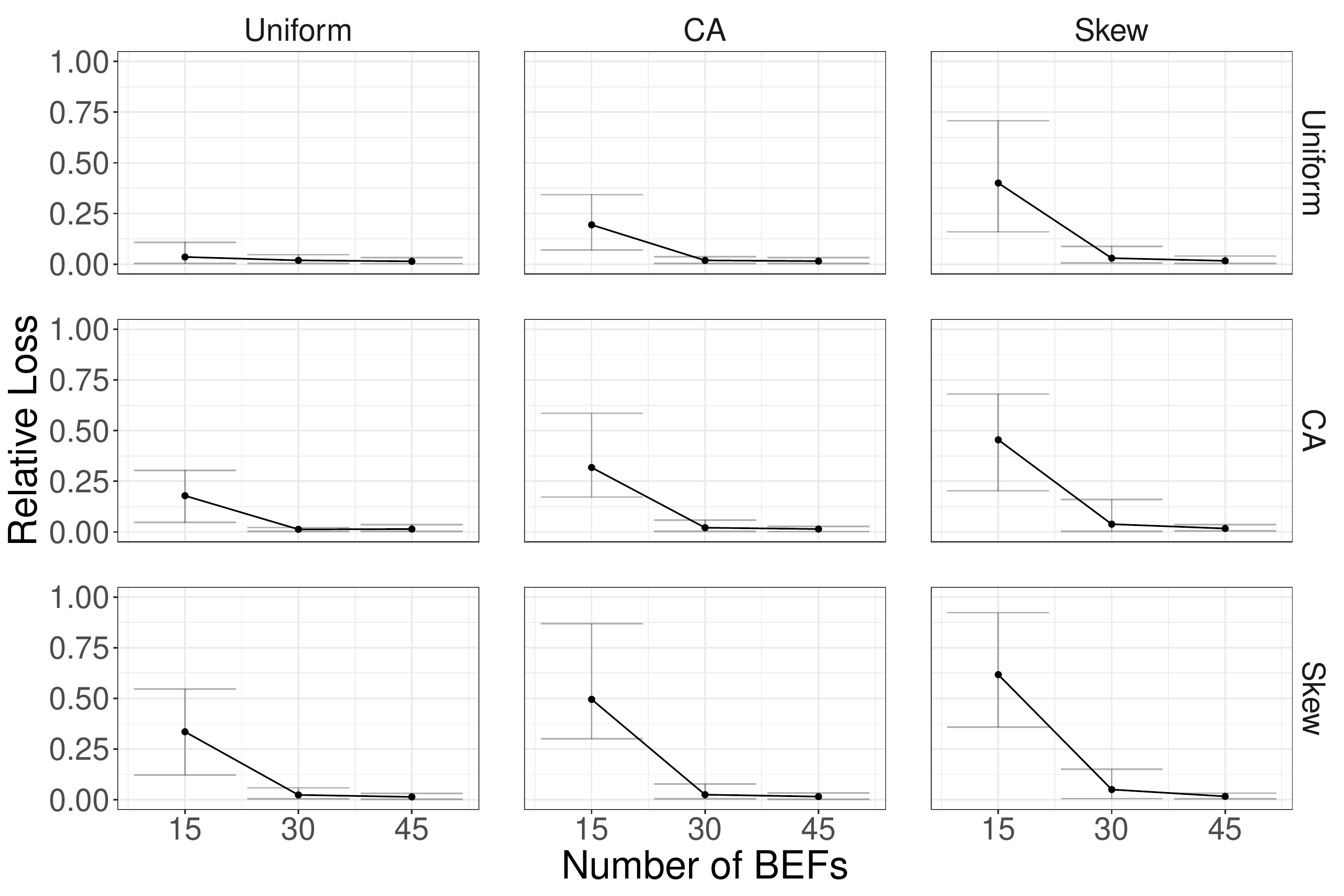}
    \caption{Relative loss as a function of different distance distributions. Points and lines represent median, 2.5 and 97.5 quantiles of loss, respectively. Row labels represent the distance distribution of the lower effect size cluster and columns that of the higher effect size cluster.}
    \label{fig:ddist_loss}
\end{figure}

Figure \ref{fig:ddist_loss} shows a number of patterns worth highlighting. 
First, across all distance distributions we observe a decrease in loss as the number of built environment features increases. 
This is as expected – more information or exposure results in a more easily detectable signal. 
Further, distance distribution combinations that include more information result in lower levels of relative loss compared to more skewed distance distributions. 
There are number of cases where this can be seen in Figure \ref{fig:ddist_loss}, the most obvious being the top-left diagonal panel where both clusters have Uniform distance distributions; this has the lowest loss values across all panels due to the relative abundance in exposure information.
This pattern holds when comparing to the more skewed distributions: The uniform-CA combination has higher error than the uniform-uniform combination when there are a relatively small number of built environment features present.

\section{Fast food restaurants near schools and child obesity among public school students in Los Angeles}
\label{sec:Results}

There is a pressing need to understand contextual determinants of child obesity, in order to implement population level strategies to reduce and prevent it\citep{IOM}. The food environment near schools has been proposed and studied as a contextual factor that influences children's diet, and thus obesity \citep{currie2010effect,davis2009proximity,sanchez2012differential,baek2016distributed}. We use data on body weight of children attending public schools in Los Angeles, CA, along with data on the locations of FFRs as a marker of the food environment near schools, and apply our proposed method to identify schools where children may be at higher risk of obesity, related to food environment exposures. Identifying these schools may help prioritize or tailor population-level interventions to address child obesity. 
\subsection{Data Description}
\label{sec:Data}
Every year public schools in the State of California collect data on the fitness status of pupils in 5th, 7th and 9th grade, as part of a state mandate, using the Fitnessgram battery of tests \citep{fgram}. 
Child-level data available for our analysis were collected during academic years 2001-2008 on 5th and 7th graders, and consists of children's weight (Kg) and height (m) and, and the following categorical covariates: sex, race-ethnicity, fitness status (unfit, fit, fit above standard), and grade level.  Weight and height are transformed to body mass index (Kg/m$^2$), and standardized to BMIz scores according to age- and sex-specific growth curves published by the United States' Center for Disease Control. In contrast to adults, standardization is needed when analyzing data from children of different ages, given children's rapid growth. 
To aid in managing the large database, and given that all child-level covariates are categorical, the dataset is ``collapsed'' so that each row represents a group of children within each school defined by the cross-classification of categorical child-level characteristics described above. The average BMIz of children in the group is the outcome of interest. Given the categorical nature of the covariates and our use of weighting by the size of the group represented by each row (below), this approach yields exactly the same results as would be obtained if data had not been grouped, thus avoiding biases in an ecological analysis \citep{schoenborn2002body}. 

\par 
Data on school-level characteristics are also available from the CDE website (see Table \ref{tbl:cluster}), and, importantly, so is the geocode of the school.
School geocodes were used for two purposes.
First, the geocodes were used to link schools to census tract level covariates. Second, the school geocodes were used to calculate the distances between the school and the geocoded location of each FFR in the LA area.
FFRs were identified from the National Establishment Time Series (NETS) database \citep{walls2013national}, using a published algorithm that classifies specific food establishments as FFRs \citep{auchincloss2012improving}. 
Only FFRs within five miles of schools were kept for this analysis. 
This distance was chosen to be a conservative as previous work estimated that the distance at which FFRs cease to have an effect on childhood obesity is approximately one mile \citep{baek2016distributed}. 

\subsection{Los Angeles STAP-DP Model}
We fit models estimating both the population-level and latent cluster-level effects – STAP and STAP-DP models, respectively. Given the available data consisting of subgroups of children defined by the cross-classification of categorical covariates, we use the standardized average BMI within the subgroup as the outcome. The models adjust for the student group and school-level covariates listed in Table \ref{tbl:cluster}.
Denoting these covariates as $\bm{X}_{ijq}$ for student group $q=1,...,n_{ij}$, measured at year $j=2001,...,2008$, attending  school $i=1,...,N$, and using notation as described in (\ref{eqn:long1}) our model for analyzing the Los Angeles data is:
\begin{align*}
 \text{BMIz}_{ijq} &= \bm{X}_{ijq}^{T}\bm{\gamma} + F_i(\mathcal{D}_{ij}) + b_{i1} + b_{i2}\frac{\text{year}_{ij}}{10} + \epsilon_{ijq},    \tag{9} \label{eqn:LA}\\  
 \epsilon_{iqj} &\sim N\left(0,\frac{\sigma^2}{n_{iqj}}\right),  \\
 \bm{b}_i &\sim MVN_2(\bm{0},\Sigma), 
 \end{align*}
 
 \noindent where $n_{ijq}$ represents the number of students in student group $q$ during year $j$ at school $i$. Given that FFRs may open or close during the study period, $\mathcal{D}_{ij}$ represents the distances between school $i$ and FFRs available within 5 miles during year $j$. 
 Similar to our simulations, we place a weakly informative Gamma(1,1) prior on each of the penalty parameters in $\tauvec$, associated with each cluster regression coefficients, the residual precision $\sigma^{-2}$, and the concentration parameter $\alpha$. 
 The Gamma(1,1) prior on the concentration parameter is a common prior setting in the DP literature, reflecting the \textit{a priori} expectation that the concentration parameter is $1$, so that fewer clusters are favored \citep{rodriguez2008nested,gelman2013bayesian}.
 Additionally, we place a non-informative Jeffrey's prior on the covariance matrix for the school specific $\bm{b}_i$ vector: $p(\Sigma^{-1}) \propto \mid \Sigma \mid^{\frac{3}{2}}$. 
 Estimation is conducted through \texttt{rstapDP}, drawing 2000 samples from each of 2 independent MCMC chains after 8000 samples have been iterated as ``burn-in'' on each chain. We check convergence via $\hat{R}$ diagnostic \cite{vehtari2020rank} and visually inspecting traceplots.
 We use $L=7$ coefficients in our spline basis function expansion, and similarly use this basis to estimate the $f_i(d)$ on a grid of values, calculating the 95\% point-wise credible interval at each distance grid point.
 We also calculate the posterior probability of co-clustering which can be arranged in a matrix $\bm{P} \in \mathbb{R}^{N\times N}$ so that $\bm{P}_{i,i'} = P($ school $i$ is co-clustered with school $i'$ across post burn-in iterations$)$. School cluster characteristics are tabulated using the cluster mode school assignment calculated using (\ref{eqn:loss}) as implemented in the \texttt{rstapDP} package via the \texttt{assign\_mode} function.
 \par 
 For comparative purposes, we fit a model similar to (\ref{eqn:LA}) in all ways save for restricting the $f_i(d)$ to be estimated at the population level - $f(d)$. We fit this model using Hamiltonian Monte Carlo via the \texttt{rsstap} R package \citep{rsstap}, drawing 1000 samples after 1000 warm-up across 4 independent MCMC chains. Convergence is assessed via $\hat{R}$ diagnostic and we calculate the analogous posterior estimate for $f(d)$ across the same grid of distance values.

\subsection{Los Angeles Results}
\begin{figure}[H]
    \centering
    \includegraphics[width = \textwidth]{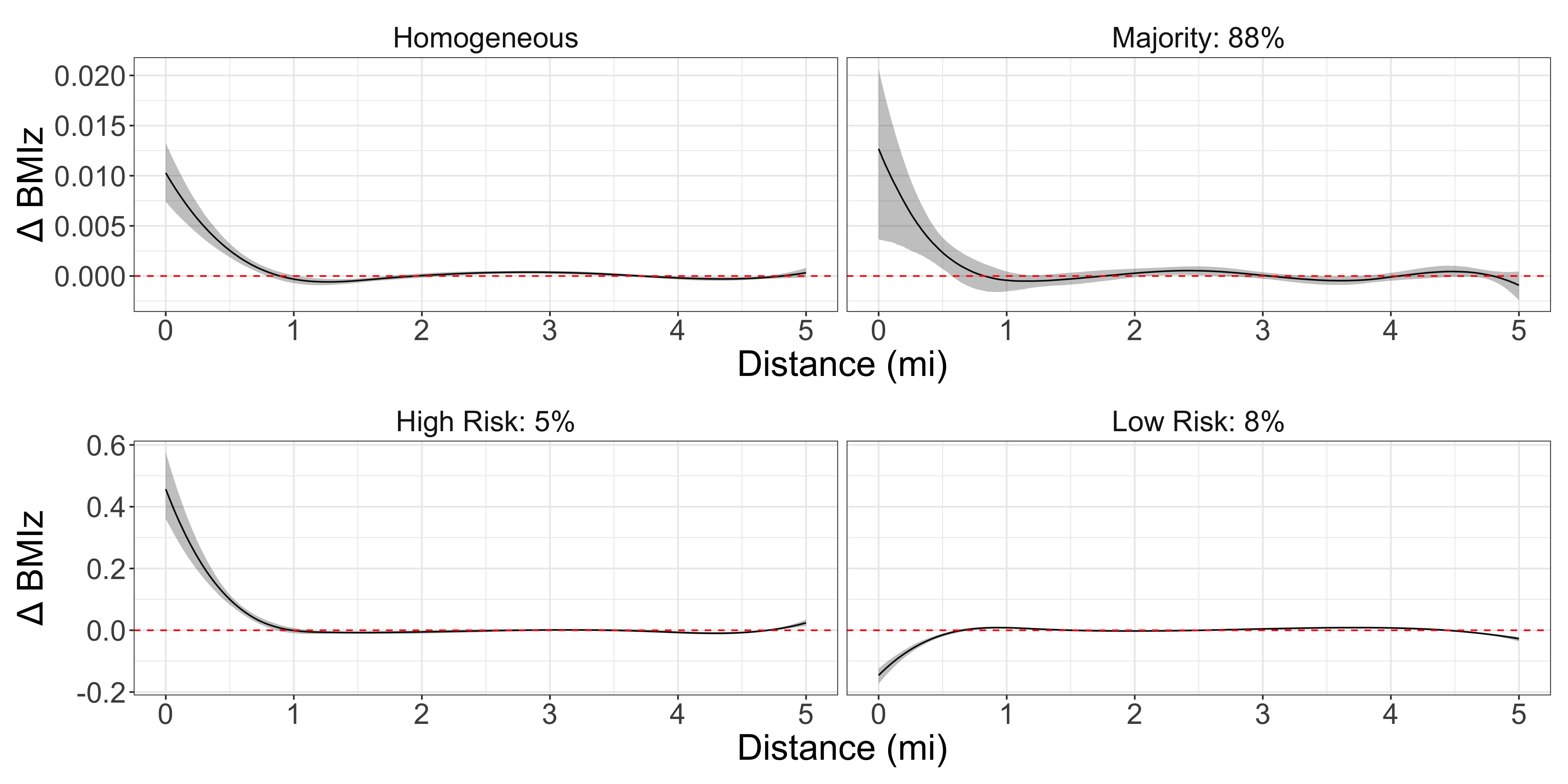}
    \caption{Changes in student BMI associated with FFR exposure across 5 mi. Line and band represents median and 95\% posterior credible interval. The number following Each cluster label represents the median proportion of schools that are assigned to that cluster. Dotted line (colored red online) represents 0 reference line. Please note that the y-axis is different between the two rows.}
    \label{fig:f_estimates}
\end{figure}

Figure \ref{fig:f_estimates} shows four functions corresponding to the 3 estimated cluster functions from the STAP-DP model, as well as the 1 homogeneous effect estimate. 
We name the three cluster effects ``Majority'', ``High Risk'' and ``Low Risk''. These names derive from the proportion of schools assigned to the cluster as well as the relative effect size associated with the function at and around distance 0 mi from the school: In the cluster labeled ``High Risk'', one additional FFR placed at distance 0 from a school is associated with an expected 0.46 higher BMIz among students attending those schools (95\% CI: 0.36, 0.58), all else equal. In contrast, placing one FFR at distance 0 from the schools assigned to the ``Low Risk'' cluster is associated with lower BMIz score, by -0.15  (95\% CI: -0.17,-0.12), all else equal. The analogous values for the ``Majority'' and homogeneous function estimates, are 0.01 (95\% CI: 0.00,0.02) and 0.01 (95\% CI: 0.01,0.0132), respectively.  In all clusters, the estimated associations rapidly decay with increasing distance, with all association estimates effectively zero by 1 mile.

We now turn our attention to the matrix of co-clustering probabilities $\bm{P}$ which we visualize using a heat-map in Figure S2 in the Supplementary material, after applying \cite{rodriguez2008nested}'s  hierarchical sorting algorithm to group schools with similar co-clustering probabilities together. 
The estimate of co-clustering probabilities  shows approximately 550 schools (90\%) are consistently co-clustered within one of the three clusters, reflecting a high degree of model certainty in cluster configurations for these schools. 
The remaining $\approx$ 50 schools show a greater uncertainty between being classified in the ``Majority'' or ``Low Risk'' cluster.
This uncertainty likely stems from an insufficient number of FFRs present within the relevant $\approx$ 1 mile distance from schools, where the cluster effects are most discernibly different.

\par Further examination of the school characteristics associated with each cluster details several suggestive, though not conclusive between-cluster differences (Table \ref{tbl:cluster}). 
 Differences between the three clusters of schools are fairly muted, with summary statistics across student- and school-level measures describing similar student populations and levels of median household income and education amongst the neighborhoods of schools in each cluster. The most noteworthy differences amongst the clusters are in the number of FFRs within $\frac{1}{2}$ mile of the school – lower in the high risk group as compared to the other two clusters – and the total enrollment – higher in the low and high risk groups as compared to the majority cluster.

\begin{table}[H]
\resizebox{\textwidth}{!}{\begin{tabular}{lrrrr}
\toprule
 & \textbf{Overall} & \textbf{Majority} & \textbf{Low Risk} & \textbf{High Risk} \\ 
  \hline
\# Students & 752,529 & 655,017 & 62,573 & 34,939 \\ 
\% Obese & 52 & 52 & 51 & 46 \\ 
Average BMIz & 0.84 & 0.85 & 0.78 & 0.71 \\ 
\% Female & 49 & 49 & 49 & 49 \\ 
 Race/Ethnicity & & & & \\
\quad   \% Asian & 3 & 2 & 2 & 7  \\ 
\quad  \% Black & 10 & 10 & 14 & 13 \\
\quad  \% Hispanic & 79 & 80 & 73 & 62 \\ 
\quad  \% White & 8 & 8 & 10 & 18 \\ 
\textbf{School Characteristic}$^{1}$ &  N = 593 &  N = 535 &  N = 36 &  N = 22\\
	\midrule
	Total Enrollment (100's of students) & 7.7 (5.0, 13.0) & 7.5 (4.9, 12.8) & 10.7 (6.3, 15.4) & 10.7 (6.2, 12.7)\\
	\# FFRs within 1/2 mile & 23 (8, 45) & 23 (8, 46) & {24} (8, 48) & 14 (5, 24)\\
	\# FFRs within 1 mile & 101 (69, 143) & 103 (68, 144) & 92 (70, 126) & 89 (74, 115)\\
	\% Free or Reduced Price Meals & 0.86 (0.69, 0.94) & 0.86 (0.69, 0.94) & 0.85 (0.75, 0.93) & 0.80 (0.58, 0.95)\\
	Education$^{2}$ & 14 (5, 27) & 14 (5, 28) & 13 (7, 25) & 13 (5, 23)\\
	Income$^{3}$ (1000 USD) & 34 (25, 49) & 34 (25, 49) & 33 (28, 52) & 32 (25, 55)\\
	School Type &  &  &  & \\ \hline
\quad 	Elementary & 468 (79\%) & 427 (80\%) & 26 (72\%) & 15 (68\%)\\
\quad 	K-12 & 7 (1.2\%) & 7 (1.3\%) & 0 (0\%) & 0 (0\%)\\
\quad 	Middle & 90 (15\%) & 75 (14\%) & 9 (25\%) & 6 (27\%)\\ 
\quad 	High School & 12 (2.0\%) & 11 (2.1\%) & 1 (2.8\%) & 0 (0\%)\\
\quad 	Other & 16 (2.7\%) & 15 (2.8\%) & 0 (0\%) & 1 (4.5\%)\\
	Urbanicity &  &  &  & \\ \hline
\quad 	Suburban & 75 (13\%) & 66 (12\%) & 6 (17\%) & 3 (14\%)\\
\quad 	Urban & 518 (87\%) & 469 (88\%) & 30 (83\%) & 19 (86\%)\\
	\bottomrule
\end{tabular}}
\caption{Characteristics of children and schools in each cluster, assigned using the mode cluster\\
\textsuperscript{1}Statistics presented: Median (IQR); n (\%) \\
\textsuperscript{2} Percent of individuals 25 years or older  within the school's census tract with at least a bachelors degree.\\
\textsuperscript{3} Median Household Income among residents in the census tract where the school is located.}
 \label{tbl:cluster}
\end{table}

\par The lack of substantial differences in measured characteristics between these two clusters is noteworthy, suggesting that  none of the observed characteristics appear to modify the obesity risk associated with FFR exposure. Although prior research shows that  area-level education and income modify the effects of child obesity interventions, for instance, the protective effects of socioeconomic factors do not appear to extend to children's risk of obesity as due to proximate FFR exposure – at least within this population. This lack of difference in socioeconomic characteristics suggests that there are unmeasured variables that account for the heterogeneous FFR effects. One potential unmeasured factor could be the type of FFRs proximal to the schools, e.g. chains vs non-chain FFR's. Another possibility could be unmeasured student-level measures of wealth –  which may modify obesity risk – but we are not able to account for in our analysis.

\section{Discussion}

\label{sec:Discussion}
This work proposed a modeling approach to identify heterogeneity in distance-dependent BEF effects. 
By allowing flexibility both across space and identifying subgroups of subjects with different effects, this modeling framework addresses two problems raised in the built environment literature, namely the MAUP and the UGCP, respectively.  The modeling approach was shown to work well in both simulated data, as well as the data that motivated this work, concerning children's BMI and proximity to FFRs near their schools. 
While spatial point pattern built environment data are the primary motivation for this methodology, it could be also be applied to temporal or spatio-temporal data, the latter which we discuss in greater detail below.
\par 
Similar to the HDLM proposed by \cite{baek2016hierarchical}, we seek to allow for differences across subjects, or other substantively defined groups like schools, in the BEF associations across space.
In contrast to that work, we pool subjects with similar association effects through the DPM, allowing us to identify latent risk subject groups.
\par 
In simulations our model demonstrated classification robustness to differing distributions of distances and expected improvement in classification due to increased information through BEF exposure or effect size. 
Our analysis of Fitnessgram data illustrated how one can analyze these data in terms of the spatial effects estimated as well as the characteristics associated with each latent cluster. 
The software to fit this model and perform the necessary auxiliary functions is freely available through our R package \texttt{rstapDP} \citep{rstapDP}.
\par 
There are a number of future directions with which to take this work. 
One obvious direction would be to extend the modeling framework for more general exponential family error distributions, though this makes estimation more difficult, as the posterior distribution of $\betvec$ is no longer available in closed form. 
Work by \cite{ferrari2020modeling} has used a Riemann Hamiltonian Monte Carlo sampler in this context for models similar to ours, without smooth functional terms. This could provide one avenue to pursue. Another direction to explore would be to incorporate temporally indexed BEF data to enable spatio-temporal function estimation via tensor product of the spline basis function expansion used here. This approach would allow for cluster estimates across space and time, increasing the dimensionality and consequently, relevancy, of this work to more precisely target and understand how environments shape health and health behaviors across both time and space.
\par 
Finally, while there have been numerous methods to identify associations between subjects and BEF exposure we believe this to be the first to utilize techniques in both the Bayesian and functional non-parametric literature to identify heterogeneous BEF effects across a population.

\section*{Acknowledgements}

This research was partially supported by NIH grants R01-HL131610 (PI: S\'anchez) and R01-HL136718 (MPIs: Sanchez-Vaznaugh and S\'anchez).\vspace*{-8pt}

\bibliographystyle{unsrtnat}
\bibliography{references.bib}

\end{document}